# Irreversible multi-band effects and Lifshitz transitions at the $LaAlO_3/SrTiO_3$ interface under field effect

*Ilaria Pallecchi [1],*, Nicolò Lorenzini [2], Mian Akif Safeen [3], Musa Mutlu Can [4], Emiliano Di Gennaro [5,6], Fabio Miletto Granozio [6], Daniele Marré [2,1]*

[1] *CNR-SPIN, c/o Dipartimento di Fisica, via Dodecaneso 33, 16146 Genova, Italy*

[2] *Università di Genova, Dipartimento di Fisica, via Dodecaneso 33, 16146 Genova, Italy*

[3] *Department of Physics, University of Poonch, Rawalakot 12350, Pakistan*

[4] *Renewable Energy and Oxide Hybrid Systems Laboratory, Department of Physics, Faculty of Science, Istanbul University, Vezneciler, Istanbul, Turkey*

[5] *Università di Napoli "Federico II", Dipartimento di Fisica "E. Pancini", Compl. Univ. di Monte S. Angelo, Via Cintia, 80126 Napoli, Italy*

[6] *CNR-SPIN, Compl. Univ. di Monte S. Angelo, Via Cintia, 80126 Napoli, Italy*

* *corresponding author e-mail: Ilaria.pallecchi@spin.cnr.it*

In this work, we investigate the irreversible effects of an applied electric field on the magnetotransport properties of $LaAlO_3/SrTiO_3$ conducting interfaces, with focus on their multiband character. We study samples of different types, namely with either crystalline or amorphous $LaAlO_3$ overlayer. Our two-band analysis highlights the similarity of the electronic properties of crystalline and amorphous interfaces, regardless much different carrier densities and mobilities. Furthermore, filling and depletion of the two bands follow very similar patterns, at least in qualitative terms, in the two types of samples. In agreement with previous works on crystalline interfaces, we observe that an irreversible charge depletion takes place after application of a first positive back gate voltage step. Such charge depletion affects much more, in relative terms, the higher and three-dimensional $d_{yz}$, $d_{zx}$ bands than the lower and bidimensional $d_{xy}$, driving the system through the Lifshitz transition from two-band to single band behavior. The quantitative analysis of experimental data evidences the roles of disorder, apparent in the depletion regime, and temperature. Noteworthy, filling and depletion of the two bands follow very similar patterns in crystalline and amorphous samples, at least in qualitative terms, regardless much different carrier densities and mobilities.

## 1. Introduction

Conducting interfaces between insulating oxides have been the object of extensive research since their discovery, [1] in view of their potential applications for multifunctional electronics, [2,3,4,5] and in particular field-effect transistors [6], quantum and superconducting devices [7] and spintronics. [8,9,10] Two remarkable features of these $SrTiO_3$-based systems are particularly worth of insight, both for the study of physical mechanisms and implications for potential applications: their multiband character and the hysteretic effects observed under field effect.

The multiband character is related to the crystal-field effect acting on the Ti $t_{2G}$ orbitals, that are split by the interface potential into $d_{xy}$ band and $d_{yz}$, $d_{zx}$ bands. The latter ones have a higher band minimum at the gamma point, form a Fermi surface with elliptical in-plane cross-section, have a more three-dimensional character, heavier in-plane electron masses and a higher density of states at the Fermi level. [11] The former one, the $d_{xy}$ band, has a lower band minimum, is more confined at the interface, has a circular in-plane cross-section of the Fermi surface, has lighter in-plane electron masses and lower density of states at the Fermi level. [11,12] Reversible band filling by electric field effect drives the system from a single band to a multiband behavior, across a Lifshitz transition, having universal character. [11] Due to the different dimensional character and spatial distribution of $d_{xy}$ band and $d_{yz}$, $d_{zx}$ bands, all the properties of the system are strongly affected by the either single or multiband character, including superconducting, [13,14] magnetic, [15,16,17] and spin-orbital [14,18] properties, thus offering the opportunity of a reversible tunability of multifunctional behavior.

The hysteretic effects are related to irreversible changes in the resistance that are systematically observed in these systems as the back gate is swept toward positive voltages at low temperatures, for the first time after cooling. [19,20,21,22,23] Such effects have been traditionally ignored in the early stages of the investigation of an electric field effect in this system. [13] The first positive back gate voltage application was applied to the samples as a "forming step", and the whole field-effect analysis was typically performed on the irreversibly created perturbed state, characterized by a lower conductivity as compared to the pristine state. Such state is metastable, but extremely persistent at low temperatures. It can be reverted to the pristine state by warming the sample to room temperature for week-long times or, as shown separately by some of the authors, [23] by visible light illumination. The connection between hysteresis in the electronic response and presence of electron traps is well-established in condensed matter science [24,25,26]. In the $LaAlO_3/SrTiO_3$ system, the observed behavior has been interpreted either in terms of Fermi level lying close to the top of the quantum well and charge leaking away from the quantum well into trap states during a positive gate polarization, also modifying the confining potential landscape [19,23] or in terms of charges trapped at oxygen vacancy sites in the $SrTiO_3$ substrate that electromigrate and cluster under the driving force of the back gate voltage. [22].

Upon these premises, many crucial issues remain still open:

1) how differently, or similarly, hysteretic effects originating in the $SrTiO_3$ substrate involve the individual $t_{2G}$ bands, i.e. the more interfacially confined $d_{xy}$ band and the more extended $d_{yz}$, $d_{zx}$ bands;

2) how hysteretic effects manifest in different kinds of samples, with either crystalline or amorphous overlayer, or with sizably different carrier density;

3) in case the existence of a “trapped charge reservoir” is confirmed, according to the hypothesis above, how does the balance of the three charge populations (in $d_{xy}$ band, in $d_{yz}$, $d_{zx}$ bands and localized in trap states) evolve, irreversibly, during a $V_G$ cycle;

4) how intrinsic inhomogeneity and disorder, particularly relevant in the depletion regime, as well as the operating temperature, affect the reversible and irreversible regimes under field effect.

In order to clarify these aspects, we deposited a crystalline and an amorphous $LaAlO_3$ films on $TiO_2$-terminated $SrTiO_3$ substrates. We carried out isothermal magnetotransport measurements under field effect at different fixed temperatures, adopting the gate voltage sweep protocol of ref. [22] and we fitted experimental data within a two-band framework, taking into account sample inhomogeneity in the depletion regime. We could thus independently monitor the populations of $d_{xy}$ and $d_{yz}$, $d_{zx}$ carriers as a function of an electric field cycle and clarify how each type of band is affected by the reversible and irreversible mechanisms induced by field effect, in both the crystalline and amorphous sample, and for different operating temperature.

## 2. Methods

### 2.1. Deposition

Crystalline and amorphous $LaAlO_3$ films were grown by Pulsed Laser Deposition (PLD) on single crystalline $TiO_2$ terminated (001) $SrTiO_3$ substrates. A KrF excimer laser ($\lambda$ = 248 nm) was used, with energy density 1.4-1.5 J/cm$^2$ and repetition rate 1Hz. In the case of crystalline $LaAlO_3$, the growth was monitored by Reflection High Energy Electron Diffraction (RHEED). The crystalline sample, called sample C1, was 10 unit cell (uc) thick. It was grown at 715 °C at an oxygen partial pressure of $10^{-3}$ mbar and post-annealed in 100mBar $O_2$ at 500°C. The amorphous $LaAlO_3$ sample, called sample A1 was ~3 nm thick. It was grown at 25 °C in $10^{-2}$ mbar oxygen partial pressure.

**2.2. Patterning**

Hall bar patterns were realized in the samples by optical lithography and low temperature (< 100 K) Reactive Ion Etching (RIE) using $BCl_3$ presented in previous papers. [27] The width and distance between voltage probes of the Hall bar were 200 μm and ~1000 μm respectively. Electrical insulation between adjacent Hall bars (> GΩ) was checked, confirming the successful outcome of the patterning process.

**2.3. Measurements**

Magnetotransport measurements were carried out in a Physical Property Measurement System (PPMS) by Quantum Design, in fields up to 9T applied in the out-of-plane direction of the samples and at temperatures down to 20 K, using the circuital configuration sketched in the inset of **Figure 3**. The gate voltage $V_G$ sweep protocol was structured as follows: $V_G$ was swept from 0V to 50V and back to 0V in steps of 25V (first voltage ramp, up to $V_{Gmax}^{(1)}$=50V), then it was swept from 0V to 100V and back to 0V in steps of 25V (second voltage ramp, up to $V_{Gmax}^{(2)}$=100V), then analogously for the voltage ramp up to $V_{Gmax}^{(3)}$= 200V, back to 0V and finally again to $V_{Gmax}^{(4)}$= 200V. A two-band analysis was performed to extract $n_1$, $n_2$, $\mu_1$, $\mu_2$ values from our curves. Details of the analysis are reported in the Supporting Information. It is worth pointing out that, in all the fitting procedures, the geometric parameter *w/d*, ratio of width to length of the conducting channel between voltage probes, was not fixed, but allowed to vary within a plausible range determined by the lithographic parameters. Indeed, *w* and *d* are not necessarily coincident with the nominal width and length, especially in the depletion regimes where the corrugation of the bottom of the conduction band related to disorder may create local variations of charge density in the submicron scale, reduced channel width and meandering current paths.

## 3. Results and discussion

The resistance curves of all the samples, shown in **Figure 1**, exhibited degenerate semiconductor temperature dependence. The Residual Resistance Ratios (*RRR20*), defined as *R(300 K)/R(20 K)*, were 88 for sample C1 and 24 for sample A1.

**Figure 2** (left panel) shows the variation of the zero-magnetic-field conductance $\sigma_{sheet}$ of sample C1 at $T = 20\ K$ as the gate voltage was swept forward and back in three successive cycles, indexed by $i = 1,2,3$. Analogous data measured at 40 K are displayed in the Supporting Information. In each cycle the voltage was ramped up to a $V_{Gmax}^{(i)}$ value (with $V_{Gmax}^{(1)}$ = 50V, $V_{Gmax}^{(2)}$ = 100V, $V_{Gmax}^{(3)}$ = 200 V, $V_{Gmax}^{(4)}$ = 200 V) and then down to zero. Clearly, reversible and irreversible regimes can be distinguished. In the increasing ramp of *1st* $V_G$ cycle (from data point 1 to 3), an anomalous regime is found, in which the conductance changes under an increasing positive field effect were negligible. In the decreasing ramps from $V_{Gmax}^{(1)}$ to zero (from data point 3 to 5), on the contrary, the conductance decreased significantly, in agreement with expectations for a $V_G$-decreasing ramp, not tracing back the previous increasing ramp. During the increasing ramps of the cycles with (i = 2,3), we observe two regimes. As long as $V_G < V_{Gmax}^{(i-1)}$ (i.e. < 50V in the second cycle and < 100 V in the third cycle), the conductance value increased, as expected under field effect, tracing back the decreasing ramp of the former cycle (data points from 5 to 7 and from 13 to 15), suggesting a regular and reversible mechanism of band filling and depletion under electric field effect. The same is observed in the final increasing ramp (data points 26 to 29). As soon as $V_G$ exceeded $V_{Gmax}^{(i-1)}$, the conductance curve switched once more to the anomalous, $V_G$-independent regime (data points from 7 to 9 and from 15 to 19).

The overall data suggest that during the anomalous $V_G$ ramps at constant conductance (data points from 1 to 3, from 7 to 9, and from 15 to 19) the sample was irreversibly, or metastably perturbed, i.e. brought to a new carrier-depleted state: the higher $V_{Gmax}^{(i)}$, the deeper the induced depletion state in the (i+1)th cycle. In the decreasing branch following $V_{Gmax}^{(3)}$=+200V, the conductance decrease exceeded one order of magnitude until, eventually, a transition to an insulating state occurred, in which the sample resistance became unmeasurably large. The transition was observed at $V_G$<50V at 20 K and at $V_G$<25V at 40 K (shown in the Supporting Information).

The right panel of Figure 2 shows the data obtained when the same cycles were performed on sample A1 at *20 K*. Analogous data measured at *10 K* are displayed in the Supporting Information. The measurements confirm the phenomenology observed on sample C1, showing no qualitative differences that could be attributed to the different nature of the samples. In quantitative terms, instead, differences were found: the conductance values were nearly one order of magnitude larger than in sample C1 and the relative conductance variations were smaller. As a result, the insulating regime was never reached, even for negative $V_G$ values applied after the last ramp to 200 V at 10 K (see inset in Figure S1 in the Supporting Information). Note that measurement temperatures were chosen for the two samples in such a way that the $V_G$ ramp protocol could be swept almost entirely without encountering the transition to the insulating state at full depletion, hence we chose *20 K* and *40 K* for the less conducting sample C1 and *10 K* and *20 K* for the more conducting sample A1. Overall, data at different temperatures indicate conductance versus $V_G$ plot exhibited qualitatively similar patterns at all temperatures, whereas conductance values increased with decreasing temperature (see Figure 2 and Figure S1 in Supporting Information). Moreover, differences were observed at *20 K* and *40 K* in sample C1 in terms of gate voltage at which the Lifshitz transition occurred, namely sample C1 reached full depletion in earlier ramps and higher $V_G$ at *20 K* as compared to *40 K*. On the contrary, in the case of sample A1, the situations at *10 K* and *20 K* were very similar both in qualitative and quantitative terms. This indicates that major temperature effects are saturated below *20 K*. The data presented in Figure 2 substantially confirm experiments that were presented in previous papers. [20,22,28]

In order to progress our understanding beyond the state of the art, we now analyze and discuss the accumulation and depletion of carriers in the $d_{xy}$ and $d_{xz,yz}$ bands under the applied field effect.

Figure 3 shows magnetoresistance and Hall resistance curves of sample C1 measured at *20 K* at different gate voltages, which qualitatively represent the behavior of all the samples. Data measured at *40 K* in this sample are shown in the Supporting Information. Curves acquired at the same gate voltage are represented by the same color, making apparent that (i) there is no unique monotonic trend of measured data as a function of $V_G$ and (ii) curves measured at the same $V_G$ after different sequences of ramp sweeps are not always bundled together. These observations confirm the existence of reversible and irreversible regimes during the voltage ramps. Magnetoresistance curves were quadratic in field for fields below 2T and then progressively tended to saturate at higher fields. In magnitude, magnetoresistance increased with increasing $V_G$ and it was larger in curves collected during increasing $V_G$ ramps in the irreversible regimes (i.e., when $V_G$ exceeded previously applied values), than in decreasing $V_G$ ramps in reversible regimes. Hall resistance curves exhibited higher slopes for low $V_G$ values, which became higher in successive decreasing ramps downward from higher $V_{Gmax}^{(i)}$ voltages. Increasing departure from linearity was seen with increasing gate voltages, more marked in the increasing $V_G$ ramps in the irreversible regimes. The corresponding magnetoresistance and Hall resistance curves of samples A1 (shown in the Supporting Information) are qualitatively similar, except for a weak localization contribution of negative magnetoresistance at low field, observed in sample A1 for negative gate voltages.

Carrying out the two-band analysis on magnetotransport data described in the Supporting Information, we could finally extract carrier densities and mobilities in the two bands, as a function of the gate voltage, in successive ramps. The results for sample C1 at 20 K are displayed in **Figure 4**. At $V_G$=0, the pristine sample exhibited two-band behavior, with $n_1$≈2.4x10$^{13}$ cm$^{-2}$ for the "lower" $d_{xy}$ band and $n_2$≈5.7x10$^{12}$ cm$^{-2}$ for the "higher" $d_{yz}$, $d_{zx}$ bands. The upper panels of Figure 4 report the total carrier density as well as the individual carrier densities in the $d_{xy}$ and in the $d_{yz}$, $d_{zx}$ bands plotted vs. $V_G$ under the same cycles discussed in Figure 2. Semitransparent segments are drawn, superposed to the data, as guides for the eyes. Large similarities with the conductance plot in the left panel of Figure 2 can be observed. In fact, anomalous ramps were observed for increasing $V_G$ (data points from 1 to 3, from 7 to 9, and from 15 to 19) for which the total carrier density was not affected by the field effect. This is surprising because, based on the nonlinear dielectric properties of $SrTiO_3$ [29], the induced electrostatic charge that is expected to be accumulated at the interface can be estimated to exceed 2.5x10$^{13}$ carriers/cm$^2$, for a $V_G$ = 200 V back gate voltage across a 0.5 mm $SrTiO_3$ substrate. During the decreasing ramps, an even more complex behavior is found. Let's consider, for example, the $n_{tot}$ curve after $V_{Gmax}^{(3)}$ = 200 V. The data initially traced back the flat curve of the increasing ramp (data points from 19 to 22) and then switched to a strongly $V_G$-dependent curve (data points from 22 to 25). In the following increasing ramp, $n_{tot}$ increased at increasing gate voltage, tracing back the decreasing ramp curve, but, once reached the initial $n_{tot}$ value, it saturated to a flat curve. The same happened in previous cycles after $V_{Gmax}^{(1)}$ and $V_{Gmax}^{(2)}$.

Confirming suggestions from previous reports, we are brought to the conclusion that some of the charges are collected in trap states, that make them unavailable for transport. Hence, under field effect cycles, three different charge reservoirs *R1*, *R2* and *R3* with populations $n_1$, $n_2$ and $n_3$ are in mutual contact with each other. The $n_1$ and $n_2$ charge carriers, respectively populating the $d_{xy}$ and the $d_{yz}$, $d_{zx}$ bands have finite $V_G$-dependent mobilities, while the *R3* charges are trapped in vanishing-mobility states and are therefore not visible as carriers in Hall measurements. The complexity of the problem is further enhanced by the fact that $n_1$, $n_2$ and $n_3$ values, rather than being determined by a ($V_G$-dependent) equilibrium condition, can remain frozen in metastable states determined by the previous history of $V_G$ cycles.

The results of our investigation can be summarized in the diagram in **Figure 5**, in which the behavior of $n_1$, $n_2$ and $n_3$ vs. $V_G$ is plotted during one and a half $V_G$ cycle, i.e. for two increasing ramps up to 200V and one decreasing ramp. The amount of total charge near the interface (trapped charge plus free charge in the quantum well) is determined by electrostatics and it depends at any time only on the $V_G$ value in that very moment, regardless of the history of $V_G$ ramps. On the contrary, the charge distribution among the three different reservoirs (the two bands and the reservoir of localized states) is history-dependent. In Figure 5, the $n_1$ and $n_2$ values represent our experimental data. The $V_G$-dependent total charge density $n_{TOT}$, indicated by the black line, was calculated on the basis of the expected *C - V* relationship in a capacitor, which is notoriously determined by geometry and dielectric constant. Knowledge of $n_{TOT}$ allows us, in turn, to estimate by difference the trapped charge $n_3$, that was assumed, for simplicity, to be zero in the pristine state at $V_G$=0. The nonlinear, T-dependent $SrTiO_3$ dielectric constant $\varepsilon_{STO}(E,T)$ was numerically calculated based on the expression reported in [29]. Qualitatively similar diagrams would describe the first two cycles up to $V_{Gmax}^{(1)}$ and $V_{Gmax}^{(2)}$. The diagram in Figure 5 suggests that the first irreversible positive ramp is characterized by an increase of the total electrostatic charge, as requested by electrostatics and charge conservation law under field effect, while the density of Hall-detectable carriers ($n_{tot}$ = $n_1$+$n_2$) remains approximately constant. This implies that the field-effect-induced excess charges fill the trapped states, only increasing the population in *R3*. During the following positive ramp, initially, from $V_G$ = 200 to 125 V, only $n_3$ decreases. Below 125 V, the depletion start affecting the Hall-detectable carriers, and especially $n_2$. At about 75 V, *R2* is completely depleted (which corresponds, experimentally, to $n_2 < 10^{12}$ $cm^{-2}$) and a Lifschitz transition to a single-band conduction takes place. We observe that, starting from an initial condition in which on *R1* and *R2* were occupied, we reached here a condition in which only *R1* and *R3* are occupied. We remark once more that the anomalous irreversible phenomena take place in the first (positive) ramp in which the system transfers mobile charge to trapped states, and these will hardly return back the charge to mobile states in next cycles. The next steps take place instead in a reversible regime. The third increasing ramp, in fact, substantially traces, back, within errors and fluctuations, the data points of the second one, in reverse order, undergoing the reverse Lifschitz transition from a single to a double conduction band. Any following decreasing/increasing ramp will bring the system to replicate the second/third ramp in Figure 5.

The Lifshitz crossover, above identified in correspondence of vanishing $n_2 << n_1$, can be also directly tracked in Figure 3 from the appearance or disappearance of non-linearity in the $R_H$ curves (see the Supporting Information for the quantitative criteria to identify the Lifshitz transitions). Such appearance or disappearance of non-linearity was found in correspondence of a lower threshold for $n_2$ around $n_2 < 3x10^{12}$ $cm^{-2}$. Altogether, six Lifshitz transitions could be identified in this way, as summarized in Table 1.

Hence in each successive ramp to increasing $V_{Gmax}^{(i)}$ the Lifshitz transition was shifted to higher voltages, because the system became increasingly and irreversibly depleted.

In the measurements on sample C1 at 40 K, a smaller non-linearity of $R_H$ curves was observed as compared to 20 K data (see Figure S2 in the supporting Information). All the gate voltages at which Lifshitz transitions occurred were shifted to higher voltages, indicating that the voltage range of single-band behavior was more extended. This must be related to the temperature dependence of the $SrTiO_3$ dielectric constant, which decreases with increasing temperature. Moreover, the Lifshitz transitions identified by the appearance or disappearance of non-linearity in the $R_H$ curves were less sharp than at 20 K. Indeed, while at 20 K a 25V $V_G$ step was enough to cause appearance/disappearance of non-linearity, at 40 K the non-linearity appeared or disappeared more gradually across multiple 25V steps. This behavior could be related to the effect of the thermal smearing of the Fermi function on the Lifshitz transition.

Similarly to carrier densities, the mobilities $\mu_1$ and $\mu_2$ of the two populations were largely unaffected in the positive voltage steps of the first cycle up to $V_{Gmax}^{(1)}$ and for every positive step in which the voltage exceeded the previous $V_{Gmax}^{(i)}$ value. Initial mobilities $\mu_1$ and $\mu_2$, shown in the lower panel of Figure 4, were 700 $cm^2$ $V^{-1}$ $s^{-1}$ and 2600 $cm^2$ $V^{-1}$ $s^{-1}$, respectively. Both $\mu_1$ and $\mu_2$ decreased with decreasing $V_G$ in successive *i-th* ramps from $V_{Gmax}^{(i)}$ to 0V. In the decreasing voltage ramp after the $V_{Gmax}^{(3)}$=200V ramp, both $\mu_1$ and $\mu_2$ changed by a factor ~3. We remind that negative (positive) gate voltage values shift the charge distributions of the two bands towards (away from) the interface, so that the mobilities of both bands may become more or less affected by interfacial defects and by electron-electron scattering.

The behavior of sample A1, though being qualitatively similar to the behavior of sample C1, shows large quantitative differences. The results of the two-band analysis at 20 K are shown in **Figure 6**. In this sample, as compared to sample C1, the total charge carrier density was one order of magnitude larger and mobilities were a factor ~3-4 smaller, $\mu_1$ in the range 150-350 $cm^2$ $V^{-1}$ $s^{-1}$ and $\mu_2$ in the range 800-1100 $cm^2$ $V^{-1}$ $s^{-1}$. The variation in relative terms of carrier density $n_2$ in the higher $d_{yz}$, $d_{zx}$ bands was much larger than the carrier density $n_1$ of the lower $d_{xy}$ band, namely from ~3x$10^{13}$ $cm^{-2}$ to full depletion. Indeed, a consistent variation that follows the same pattern as in sample C1 was observed only for $n_2$. This appears plausible, because $n_2$ carriers are further away from the interface than *n1* carriers and on average higher in energy, hence more sensitive to the back gate voltage. Moreover, unlike for the low-carrier-density C1 sample, the initial $n_2$ carrier density was exactly comparable to the density of carriers that can be displaced by field effect in our back gate geometry, taking into account the non-linearity of the $SrTiO_3$ dielectric constant. As a result, the variation of $n_2$ compensated for the variation in polarization charge on the insulating $SrTiO_3$ barrier, shielding $n_1$ from any effect. Also $\mu_1$ and $\mu_2$ remained on average unchanged. It is interesting to note that a single-band analysis (not shown), the variation of $n_2$ was directly reflected in an apparent change in mobility, because the conductance expressed in terms of two-band parameters is $\sigma_{sheet}=q(n_1\mu_1+n_2\mu_2)$, so that the single band mobility $\mu=(n_1\mu_1+n_2\mu_2)/n$ is affected by the variation of $n_2$, which is multiplied by the higher mobility $\mu_2$, while the total carrier density $n$ remains practically constant, with only small relative variations.

Regarding Lifshitz transitions in sample A1 at 20 K, again assuming around $n_2 < 3\text{x}10^{12}$ $cm^{-2}$ as a threshold, full depletion was reached at $V_G$=0V in the decreasing branch of the ramp to 100 V, and at $V_G$=100V in the decreasing branch of the ramp to 200 V. Depletion required stronger "forming steps" in sample A1 as compared to sample C1, due to the large difference of total charge density and mobilities in these samples.

Finally, regarding the role of disorder, we note that in the fitting results, *w/d* changed by a factor up to 1.5-2 from the accumulation to the deep depletion regimes in both samples, indicating that when the Fermi level approached the corrugated bottom of the conduction band, patches of insulating areas appeared along the current path, this affecting the current path length d and width w. This evidences the role of disorder in these systems.

Further dedicated experiments will be needed to assess the ultimate atomic origin of the set of localized states forming the third charge reservoir. Some suggestions about the origin of such traps has been proposed in previous papers. According to some authors [19,20,23], the mechanism of charge trapping is purely electronic, assuming a Fermi level lying close to the top of the quantum well and charge carriers thermally escaping from the quantum well into traps deep in the $SrTiO_3$ substrate, during a positive gate polarization. In ref. [22], an alternative ionic mechanism was proposed, in terms of oxygen vacancy electromigration and clustering in the $SrTiO_3$ substrate under the driving force of the back gate voltage. Such oxygen vacancy clusters would form in-gap states that trap the free charge of the quantum well during the first positive gate polarization and do not release it as long as the sample is kept at low temperature. To our understanding, neither of the two hypotheses is supported so far by conclusive evidence.

**4. Conclusions**

We performed magnetotransport measurements under field effect in crystalline and amorphous $LaAlO_3/SrTiO_3$ interfaces, addressing the irreversible behavior emerging when the mobile charge density is plotted vs. $V_G$. We found that in both types of interfaces a two-band model describes data quantitatively. We postulated the existence of an extra net surface density of trapped charge, invisible to transport measurements, that allows to properly account, given the gate voltage and device capacitance, for the total field-induced surface charge variation.

Based on such approach, the apparent inconsistency of the $V_G$-dependent magnetotransport results discussed in previous papers is interpreted in terms of a three-reservoir model, in which electrons can potentially occupy either two different types of band states ((1) $d_{xy}$ type or (2) $d_{yz}$, $d_{zx}$ type), or (3) localized states, rearranging their populations under field effect. The presence of trap (localized) states induces an irreversible component in the electronic response to a $V_G$ cycle. In fact, while the total electronic charge at the interface is determined at any time by the actual $V_G$ value, the charge distribution between the three reservoirs is history-dependent. Purely electronic, [20,23] purely ionic, or mixed [22] mechanisms could be responsible for the trapping of electrons in the localized states defined in this work as "third charge reservoir", and such trapping ultimately determines the hysteretic behavior. The data as well as the phenomenological phase diagram reported in this work leave the nature and origin of the trapped charge, so far, unaddressed.

Our results show that the $d_{xy}$ and the $d_{yz}$, $d_{zx}$ bands, having a different dimensionality, a different spatial distribution and a different minimum energy at the gamma point, are differently affected by the $V_G$ variation. The spatially more extended carriers in $d_{yz}$, $d_{zx}$ bands are mostly affected and, as a result, partially screen the deeper (in energy) and more interfacial $d_{xy}$ carriers. Accordingly, $d_{yz}$, $d_{zx}$ bands are more prone to getting irreversibly depleted by exchanging their charge with the localized, "trapping" states reservoir.

In spite of qualitative similarities, quantitatively different effects were found in the crystalline and amorphous sample, which can be largely attributed to their different carrier densities. In particular, in the high-carrier-density amorphous sample, the $V_G$-induced $d_{yz}$, $d_{zx}$ carrier variations were able to shield almost completely the carriers from $d_{xy}$ bands, which in turn remained largely unaffected. In both samples a full irreversible depletion of $d_{yz}$, $d_{zx}$ band was achieved under appropriate $V_G$ cycles and in the low- carrier-density crystalline sample a full depletion of $d_{xy}$ band was achieved as well, proving the possibility to engineer irreversible Lifshitz transitions at different gate voltages.

Finally, we addressed the roles of disorder and temperature. We found that ever present inhomogeneity of the samples must be taken into account in the data analysis, particularly in the depletion regime, when the Fermi level approaches the corrugated bottom of the conduction band, and this makes geometric parameters (width to length ratio) of the conducting channel unknown. We evidenced the role of the operating temperature, which was visible much more in the 20 K-40 K range than in the 10 K-20 K range. The effect of temperature was visible via the temperature dependence of the $SrTiO_3$ dielectric constant, relevant in determining electric field values at given gate voltages, and via the smearing of the Fermi function, relevant as the Fermi level crosses the Lifshitz transition.

**Acknowledgements**

The authors acknowledge funding from the projects QUANTOX (QUANtum Technologies with 2D-OXides) of QuantERA ERA-NET Cofund in Quantum Technologies (Grant Agreement N. 731473) implemented within 10th European Unions Horizon 2020 Programme, MIUR PRIN 2017 (Grant No. PRIN 20177SL7HC TOP-SPIN), and Università di Genova through the "Fund for promoting European projects"

**Conflict of Interest**

The authors declare no conflict of interest.

| →→→→ successive ramps →→→→ | | | | | | |
|---|---|---|---|---|---|---|
| **From single band to two-band** | | $V_G$=25V in the increasing ramp to $V_{Gmax}^{(2)}$=100V (**6**) | | $V_G$=50V in the increasing ramp to $V_{Gmax}^{(3)}$=200V (**14**) | | $V_G$=100V in the final increasing ramp to $V_{Gmax}^{(4)}$=200V (**27**) |
| **From two-band to single band** | $V_G$=0 in the decreasing ramp from $V_{Gmax}^{(1)}$=50V (**5**) | | $V_G$=25V in the decreasing ramp from $V_{Gmax}^{(2)}$=100V (**12**) | | $V_G$=100V in the decreasing ramp from $V_{Gmax}^{(3)}$=200V (**23**) | |

**Table 1**. Lifshitz transitions from single band to two-band behavior and viceversa, identified in sample C1 in successive ramps of $V_G$ at 20 K. Columns from left to right correspond to successive stages in the $V_G$ ramp protocol. Numerical labels in bold font in round parentheses indicate the sequential order of data point measurement as in Figure 2.

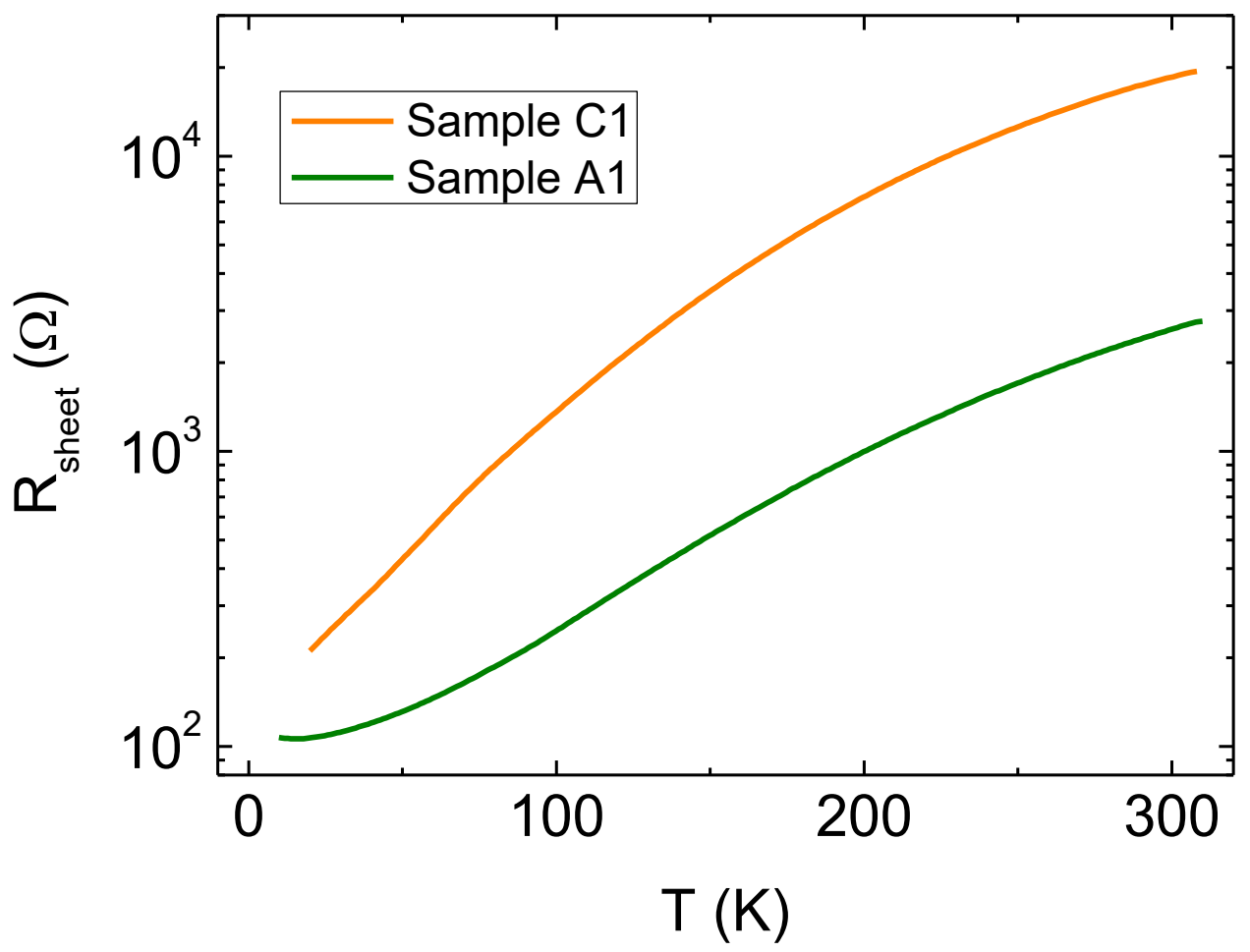

**Figure 1:** Resistance versus temperature curves of samples C1 and A1.

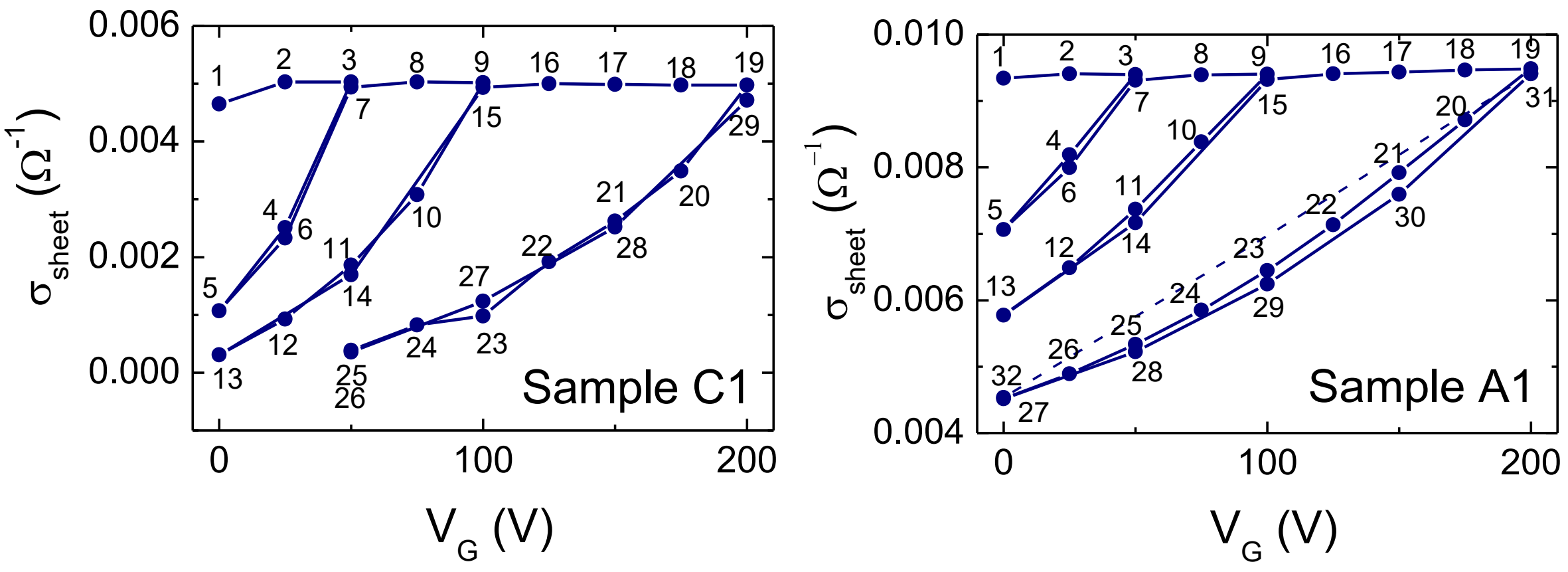

**Figure 2:** Zero field longitudinal conductance $\sigma_{sheet}$ of samples C1 (left) and A1 (right) as a function of applied gate voltage $V_G$, in successive $V_G$ ramps from zero to $V_{Gmax}^{(i)}$, measured at 20 K. The continuous lines between data points and the numerical labels indicate the sequential order of data point measurement. The last point $V_G$=0V (32) was taken to confirm once more the reproducibility of repeated cycles. We used a dashed line, to avoid confusions, to connect it to the previous data point $V_G$=200V (31).

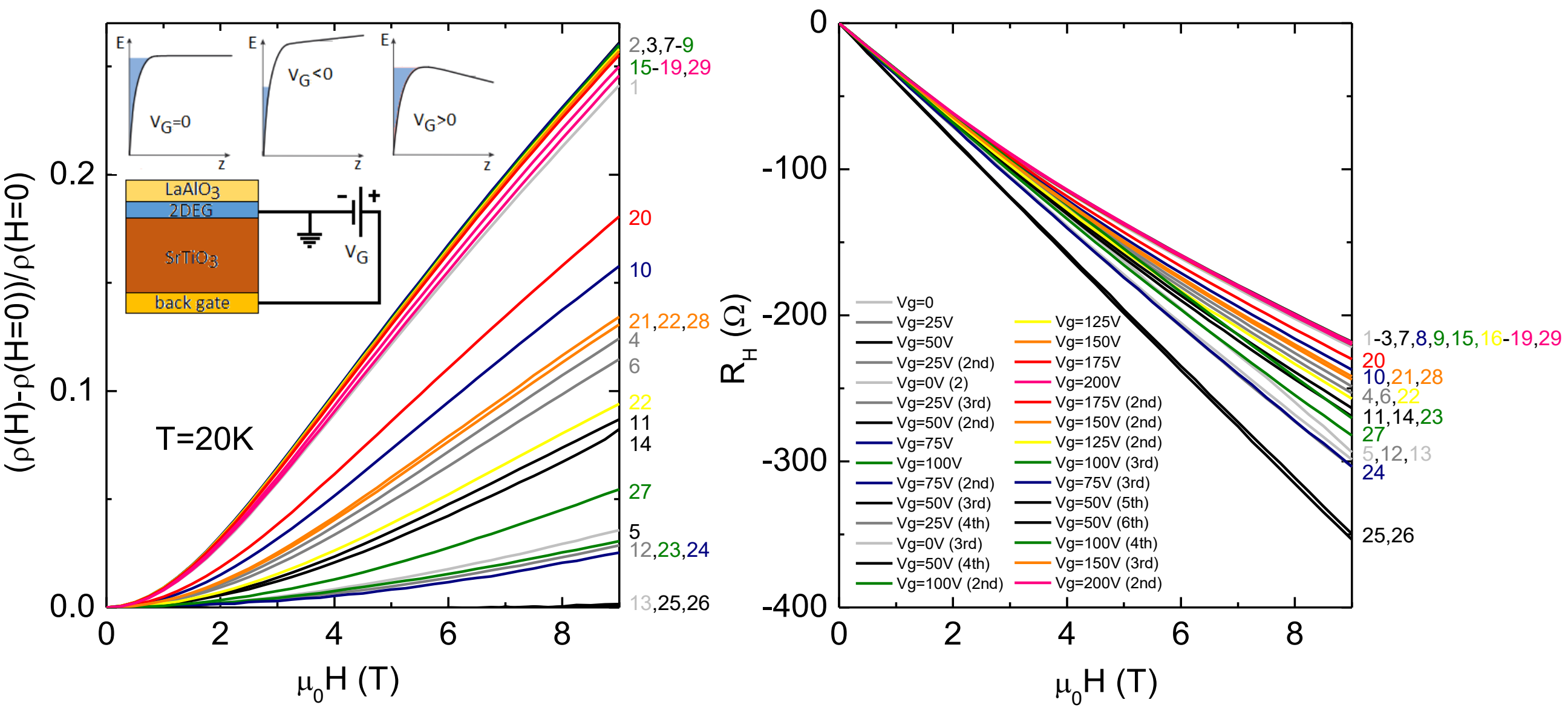

**Figure 3:** Longitudinal magnetoresistance (upper panel) and Hall resistance (lower panel) of sample C1 at 20 K, at different gate voltages. Curves that share the same color are acquired at the same gate voltage. Numeric labels on the right-hand side of the plots, having the same color code as the corresponding experimental curves, indicate the sequential order of the applied voltage. In the inset, the two dimensional electron gas (2DEG) at the interface and the circuital configuration are sketched.

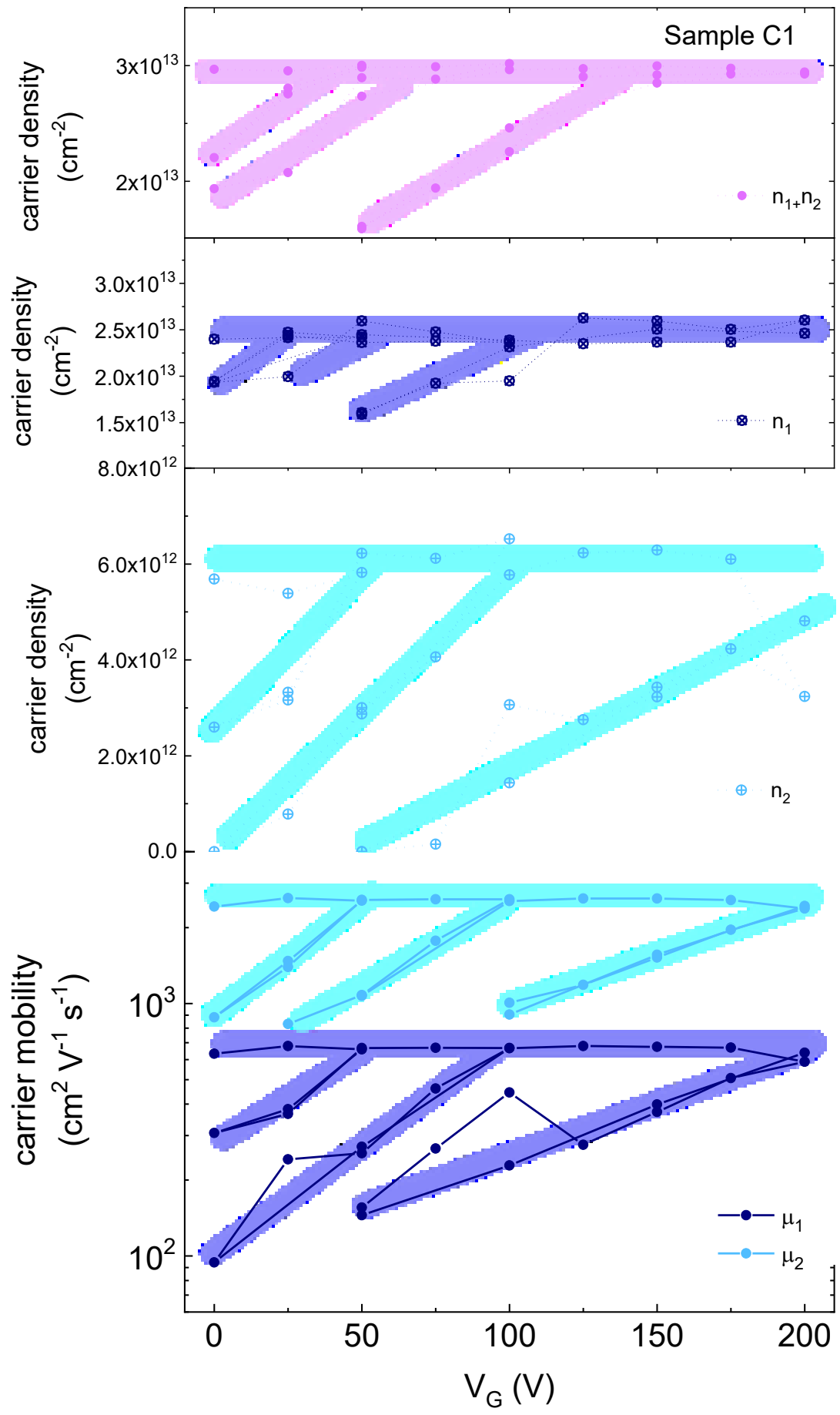

**Figure 4:** Densities (upper panels) and mobilities (lower panel) of charge carriers in the two bands in sample C1 at 20 K.

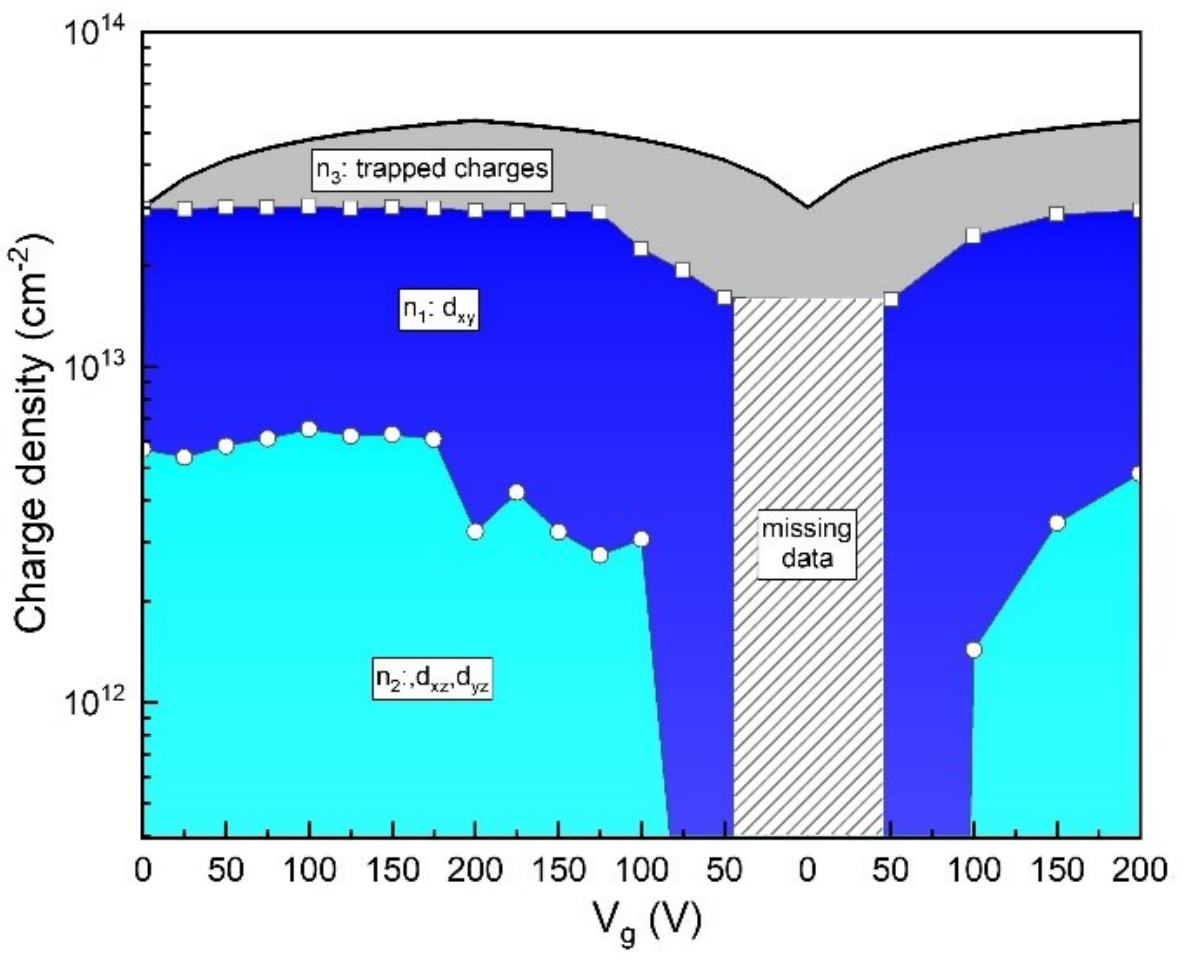

**Fig. 5:** Schematic diagram of charge densities in the two bands and in localized trap states during two increasing ramps up to 200V and one decreasing ramp. The $n_1$ and $n_2$ values represent our experimental data. The total charge, delimited by the black line, includes a calculated $V_G$-dependent carrier density, allowing to deduce by difference the amount of trapped charge $n_3$ as a function of the gate voltage. For simplicity, $n_3$ is assumed to be zero in the pristine state at $V_G$=0.

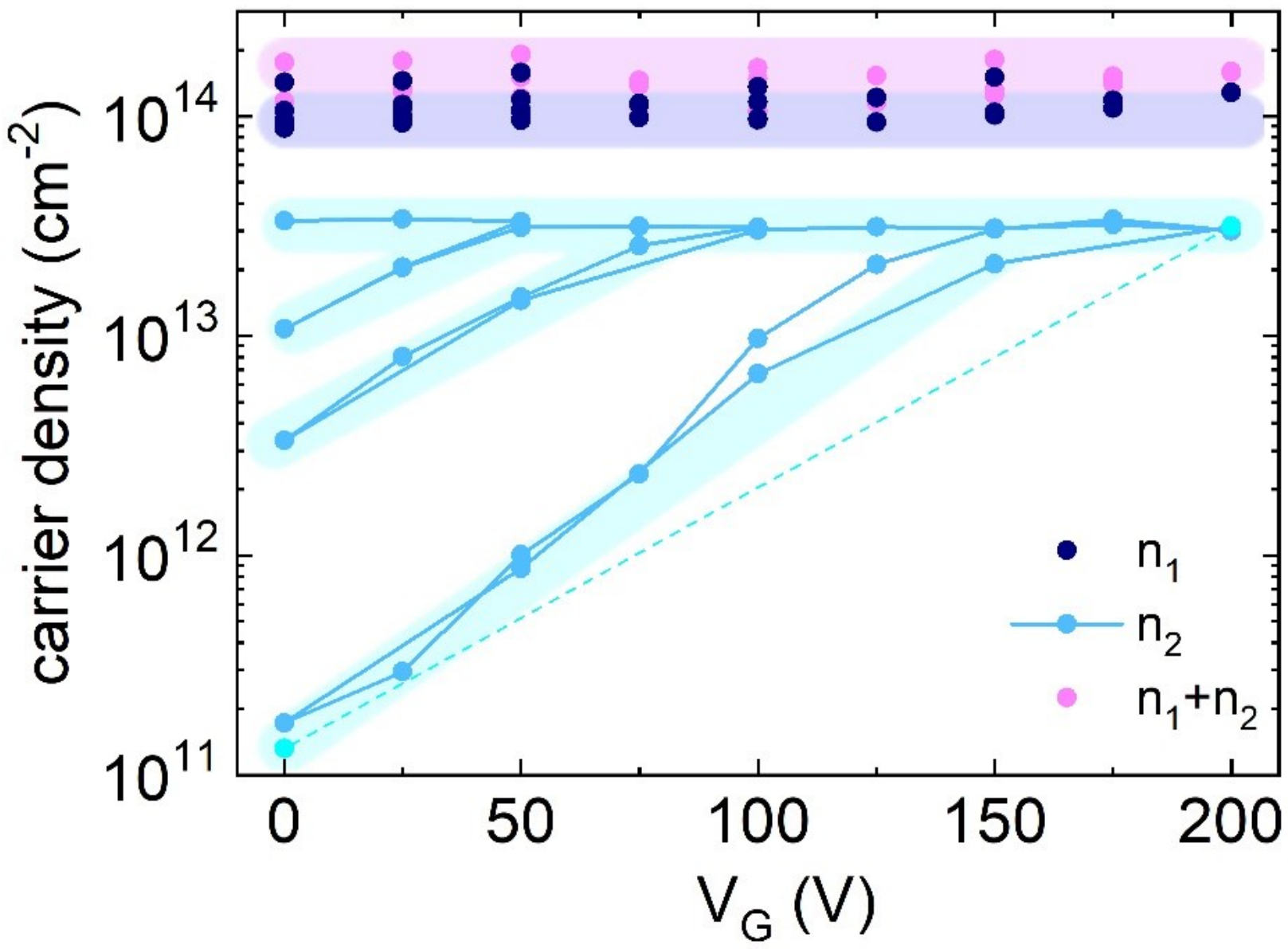

**Figure 6:** Band-resolved charge carrier densities in sample A1 at 20 K. The continuous lines between $n_2$ data points are reported to allow tracing the sequential order of data point measurement, the same as in Fig. 2b. The last point $V_G$=0V (32) was taken to confirm once more the reproducibility of repeated cycles. We used a dashed line, to avoid confusions, to connect it to the previous data point $V_G$=200V (31).

**Table of contents**

*Ilaria Pallecchi *, N. Lorenzini, Mian Akif Safeen, Musa Mutlu Can, E. Di Gennaro, F. Miletto Granozio, D. Marré*

**Irreversible multi-band effects and Lifshitz transitions at the LaAlO3/SrTiO3 interface under field effect**

Reversible and irreversible effects of an electric field on the magnetotransport properties of $LaAlO_3/SrTiO_3$ interfaces are explored within a multiband framework. Irreversible charge depletion takes place after first application of a positive back gate voltage and it involves much more, in relative terms, the higher and three-dimensional $d_{yz}$, $d_{zx}$ bands than the lower and bidimensional $d_{xy}$, driving the system through a Lifshitz transition.

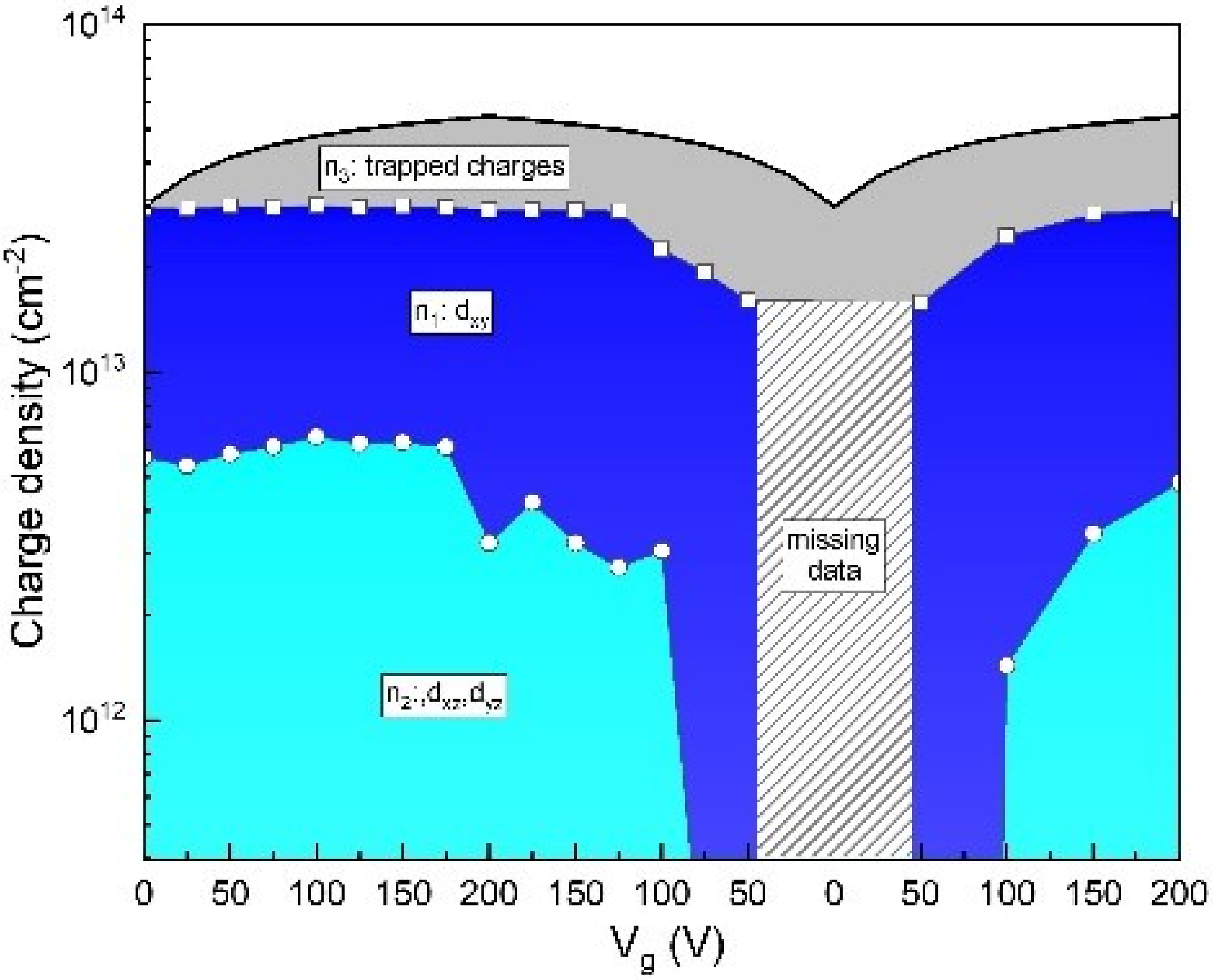

ToC figure ((Please choose one size: 55 mm broad × 50 mm high **or** 110 mm broad × 20 mm high. Please do not use any other dimensions))

[27] M. Z. Minhas, H. H. Blaschek, F. Heyroth, G. Schmidt, "Sidewall depletion in nano-patterned LAO/STO heterostructures", **2016**, *AIP Advances* 6(3) 5002.

[28] M. Zwiebler, E. Di Gennaro, J. E. Hamann-Borrero, T. Ritschel, R. J. Green, G. A. Sawatzky, E. Schierle, E. Weschke, A. Leo, F. Miletto Granozio & J. Geck, "Transition from a uni- to a bimodal interfacial charge distribution in $LaAlO_3/SrTiO_3$ upon cooling", **2020**, *Sci. Rep.* 10, 18359.

[29] S. Hirose, H. Okushi, S. Ueda, H. Yoshikawa, Y. Adachi, A. Ando, T. Ohsawa, H. Haneda, N. Ohashi, "Electric field and temperature dependence of dielectric permittivity in strontium titanate investigated by a photoemission study on $Pt/SrTiO_3$:Nb junctions", **2015**, *Appl. Phys. Lett*. 106, 191602.